\documentstyle[12pt,epsf]{article}

\newcommand{\be}{\begin{equation}}
\newcommand{\ee}{\end{equation}}
\newcommand{\bea}{\begin{eqnarray}}
\newcommand{\eea}{\end{eqnarray}}

\newcommand{\aeq}{&=&}

\newcommand{\itDelta}{{\it \Delta}}

\newcommand{\itPi}{{\it \Pi}}

\newcommand{\bra}{\langle}
\newcommand{\ket}{\rangle}
\newcommand{\dbra}{\bra \! \bra}
\newcommand{\dket}{\ket \! \ket}

\newcommand{\bq}{\bar q}

\newcommand{\rS}{{\rm S}}

\begin{document}
\addtolength{\abovedisplayskip}{5pt}
\addtolength{\belowdisplayskip}{5pt}
\setlength{\baselineskip}{20pt}

\title{An Aspect of Granulence in view of Multifractal Analysis}
\author{Naoko Arimitsu\\
Department of Computer Engineering, Yokohama National University\\ 
Kanagawa 240-8501, Japan\\
and\\
Toshihico Arimitsu\\
Institute of Physics, University of Tsukuba\\
Ibaraki 305-8571, Japan}

\date{\today}

\maketitle

\begin{abstract}
The probability density function of velocity fluctuations of {\em glanulence} 
observed by Radjai and Roux in their two-dimensional simulation of 
a slow granular flow under homogeneous quasistatic shearing 
is studied by the multifractal analysis for fluid turbulence 
proposed by the present authors.
It is shown that the system of granulence and of turbulence have indeed 
common scaling characteristics.
\\
\noindent
{\bf Keywords:} multifractal analysis, velocity fluctuation, turbulence, granulence
\end{abstract}

\section{Introduction}

In this paper, we apply the multifractal analysis (MFA) 
\cite{AA,AA1,AA10,AA12}
of fluid turbulence to granular turbulence ({\em granulence} \cite{Radjai02})
in order to see how far MFA works in the study of the data
observed by Radjai and Roux \cite{Radjai02} in their two-dimensional simulation 
of a slow granular flow subject to homogeneous quasistatic shearing. 
Radjai and Roux reported that there is an evident analogy between
the scaling features of turbulence and of granulence in spite of the fundamentally
different origins of fluctuations in these systems.
MFA is a unified self-consistent approach for the systems with large deviations,
which has been constructed based on 
the Tsallis-type distribution function \cite{Tsallis88}
that provides an extremum of the {\it extensive} R\'{e}ny \cite{Renyi} 
or the {\it non-extensive} Tsallis entropy \cite{Tsallis88,Havrda-Charvat}
under appropriate constraints.

\section{Multifractal Analysis}

MFA of turbulence rests
on the scale invariance of the Navier-Stokes equation for high Reynolds number, 
and on the assumption that the singularities due to the invariance 
distribute themselves multifractally in physical space.

The velocity fluctuation 
$
\delta u_n = \vert u(\bullet + \ell_n) - u(\bullet) \vert
$
of the $n$th multifractal step satisfies the scaling law
$
\vert u_n \vert \equiv \left\vert \delta u_n/\delta u_0 \right\vert 
= \delta_n^{\ \alpha/3}
$
with
$
\delta_n = \ell_n /\ell_0 = \delta^{-n}
\label{r-n}
$
$
(n=0,1,2,\cdots)
$.
We call $n$ the multifractal depth which can be real number in the analysis of 
experimental data.
We will put $\delta =2$ in the following in this paper that is consistent with
the energy cascade model.
At each step of the cascade, say at the $n$th step, eddies break up into 
two pieces producing the energy cascade with the energy-transfer rate
$\epsilon_n$ that represents the rate of transfer of energy per unit mass 
from eddies with diameter $\ell_n$ to those with $\ell_{n+1}$.
Then, we see that the velocity derivative 
$
\vert u^\prime \vert = \lim_{n \rightarrow \infty} u^\prime_n
$
with the $n$th velocity difference
$
u'_n = \delta u_n/\ell_n
$
for the characteristic length $\ell_n$
diverges for $\alpha < 3$.
The real quantity $\alpha$ is introduced 
in the scale transformation \cite{Frisch-Parisi83,Meneveau87b}
$
{\vec x} \rightarrow {\vec x}'=\lambda {\vec x},\ 
{\vec u} \rightarrow {\vec u}'=\lambda^{\alpha/3} {\vec u},\ 
t \rightarrow t'=\lambda^{1- \alpha/3} t,\ 
p \rightarrow p'=\lambda^{2\alpha/3} p
\label{scale trans}
$
that leaves the Navier-Stokes equation 
$
\partial {\vec u}/\partial t
+ ( {\vec u}\cdot {\vec \nabla} ) {\vec u} 
= - {\vec \nabla} p
+ \nu \nabla^2 {\vec u}
\label{N-S eq}
$
of incompressible fluid invariant for a large Reynolds number 
$
{\rm Re}=\delta u_{\rm in} \ell_{\rm in}/\nu
$.
Here, $\nu$ is the kinematic viscosity, $p=\check{p}/\rho$ with
the thermodynamical pressure $\check{p}$ and the mass density $\rho$, and
$\delta u_{\rm in}$ and $\ell_{\rm in}$ represent, respectively, 
the rotating velocity and the diameter of the largest eddies in turbulence.
The largest size of eddies is, for example, about the order of mesh size of the grid, 
inserted in a laminar flow, which produces turbulence downstream.

Within MFA, it is assumed that the singularities due to the scale invariance 
distribute themselves, multifractally, in physical space with 
the Tsallis-type distribution function, i.e.,
the probability 
$
P^{(n)}(\alpha) d\alpha
$ 
to find in real space a singularity with the strength $\alpha$ within the range
$
\alpha \sim \alpha + d \alpha
$
is given by \cite{AA1,AA3,AA4}
$
P^{(n)}(\alpha) = ( Z_{\alpha}^{(n)} )^{-1} \{ 1  - 
[ \left(\alpha - \alpha_0\right)/\itDelta \alpha ]^2 \}^{n/(1-q)}
\label{Tsallis prob density}
$
with 
$
(\itDelta \alpha)^2 = 2X/[(1-q) \ln 2]
$.
Here, $q$ is the entropy index introduced in the definitions of 
the R\'enyi and the Tsallis entropies.
This distribution function provides us with the multifractal spectrum
$
f(\alpha) = 1 + (1-q)^{-1} \log_2 [ 1 - 
(\alpha - \alpha_0 )^2/(\Delta \alpha )^2 ]
\label{Tsallis f-alpha}
$
which, then, produces the mass exponent
\bea
\tau(\bq) = 1-\alpha_0 \bq + 2X\bq^2 (1+ \sqrt{C_{\bq}})^{-1}
+ (1-q)^{-1} [1-\log_2 (1+ \sqrt{C_{\bq}} ) ]
\label{tau}
\eea
with
$
{C}_{\bq}= 1 + 2 \bq^2 (1-q) X \ln 2
\label{cal D}
$.
The multifractal spectrum and the mass exponent are related with each other through 
the Legendre transformation \cite{Meneveau87b}:
$
f(\alpha) = \alpha \bq + \tau(\bq)
\label{f-tau1}
$
with
$
\alpha = - d \tau(\bq)/d \bq
\label{alpha-tau1}
$
and
$
\bq = d f(\alpha)/d \alpha
\label{def of bq1}
$.

The formula of the probability density function (PDF)
$
\Pi^{(n)}(u_n) 
$
of velocity fluctuations is assumed to consists of two parts, i.e.,
$
\Pi^{(n)}(u_n) = \Pi^{(n)}_{\rS}(u_n) + \Delta \Pi^{(n)}(u_n) 
\label{def of Pi}
$
where the first term is related to $P^{(n)}(\alpha)$ by
$
\Pi^{(n)}_{\rS}(\vert u_n \vert) du_n \propto P^{(n)}(\alpha) d \alpha
\label{singular portion}
$
with the transformation of the variables 
$
\vert u_n \vert = \delta_n^{\ \alpha/3}
$, 
and the second term is responsible to the contributions coming from 
the dissipative term in the Navier-Stokes equation violating the invariance under 
the scale transformation given above.
Then, we have the velocity structure function in the form
$
\dbra \vert u_n \vert^m \dket \equiv \int du_n  
\vert u_n \vert^m \Pi^{(n)}(u_n)
= 2 \gamma^{(n)}_{m}
+ (1-2\gamma^{(n)}_{0} ) a_{m} \delta_n^{\ \zeta_{m}}
\label{structure func m}
$
with
$
2\gamma^{(n)}_{m} = \int du_n 
\vert u_n \vert^m \Delta \Pi^{(n)}(u_n)
$,
$
a_{m} = \{ 2/[\sqrt{C_{m/3}} ( 1+ \sqrt{C_{m/3}})] \}^{1/2}
$
and the scaling exponent
$
\zeta_m = 1-\tau(m/3)
$
given with the mass exponent (\ref{tau}).

The PDF 
$
\hat{\Pi}^{(n)}(\xi_n)
$
both of velocity fluctuations and of velocity derivative 
to be compared with observed data is the one defined through
$
\hat{\Pi}^{(n)}(\xi_n) d\xi_n = \Pi^{(n)}(u_n) d u_n
$
with the variable 
$
\xi_n = u_n/\dbra u_n^2 \dket^{1/2}
$
scaled by the standard deviation of velocity fluctuations.
For the velocity fluctuations larger than the order of its standard deviation, 
$\xi_n^* \leq \vert \xi_n \vert$ (equivalently, $\vert \alpha \vert \leq \alpha^*$), 
the PDF is given by \cite{AA10,AA12}
\bea
\hat{\Pi}^{(n)}(\xi_n) d \xi_n 
\aeq \Pi^{(n)}_{\rm S} (u_n) du_n
\nonumber\\
\aeq \bar{\Pi}^{(n)} \frac{\bar{\xi}_n}{\vert \xi_n \vert}
\left[1 - \frac{1-q}{n}\ 
\frac{\left(3 \ln \vert \xi_n / \xi_{n,0} \vert\right)^2}{
2 X \vert \ln \delta_n \vert} \right]^{n/(1-q)} d \xi_n
\label{PDF large}
\eea
with
$
\xi_{n,0} = \bar{\xi}_n \delta_n^{\alpha_0 /3 -\zeta_{2} /2}
$ and
$
\bar{\Pi}^{(n)}  
= 3 (1-2\gamma^{(n)}_0)/(2 \bar{\xi}_n 
\sqrt{2\pi X \vert \ln \delta_n \vert})
$.
This {\em tail part} represents the large deviations, and manifests itself 
the multifractal distribution of the singularities due to the scale invariance 
of the Navier-Stokes equation when its dissipative term can be neglected.
The entropy index $q$ should be unique once a turbulent system with 
a certain Reynolds number is settled.
For smaller velocity fluctuations, 
$
\vert \xi_n \vert \leq \xi_n^*
$ (equivalently, $\alpha^* \leq \vert \alpha \vert$),
we assume the Tsallis-type PDF of the form \cite{AA10,AA12}
\bea
\lefteqn{\hat{\Pi}^{(n)}(\xi_n) d \xi_n =
\left[ \hat{\Pi}^{(n)}_{\rS}(u_n)
+\Delta \hat{\Pi}^{(n)}(u_n) \right] d u_n}
\nonumber\\
\aeq \bar{\Pi}^{(n)}
\left\{1-\frac{1-q'}{2} \left(1+3f'(\alpha^*)\right)
\left[ \left(\frac{\xi_n}{\xi_n^*}\right)^2 -1 \right] \right\}^{1/(1-q')} d \xi_n
\label{PDF small}
\eea
where a new entropy index $q'$ is introduced as an adjustable parameter.
This {\em center part} is responsible to smaller fluctuations, compared with 
its standard deviation, due to the dissipative term violating the scale invariance.
The entropy index $q'$ can be dependent on the distance of two measuring points.

The two parts of the PDF, (\ref{PDF large}) and (\ref{PDF small}), are
connected at 
$
\xi_n^* = \bar{\xi}_n \delta_n^{\alpha^* /3 -\zeta_{2} /2}
$
with the conditions that they have a common value and that their slopes coincide.
The value $\alpha^*$ is the smaller solution of
$
\zeta_{2}/2 - \alpha/3 +1 -f(\alpha) = 0
$.
The point $\xi_n^*$ has the characteristics that the dependence of
$
\hat{\Pi}^{(n)}(\xi_n^*)
$
on $n$ is minimum for $n \gg 1$.
With the help of the second equality in (\ref{PDF small}) and (\ref{PDF large}),
we obtain $\Delta \Pi^{(n)}(x_n)$, and have the analytical formula 
to evaluate $\gamma_{m}^{(n)}$. Their explicit analytical formulae and 
the definition of 
$
\bar{\xi}_n
$
are found in \cite{AA10,AA12}.

\section{Turbulence}

\begin{figure}[htb]
\begin{center}
\leavevmode
\epsfxsize=120mm
\epsfbox{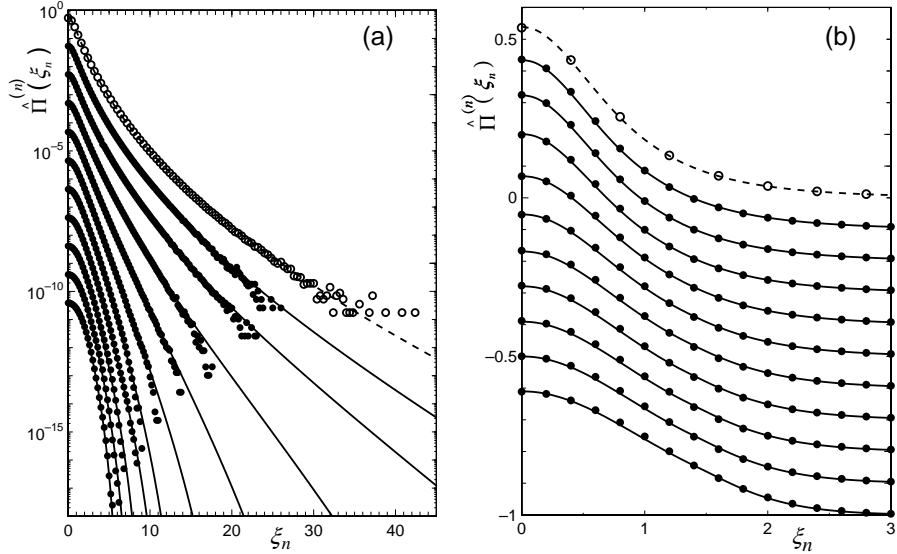}
\label{velocity fluctuations log}
\caption{{\small
Analyses of the PDF's of {\it velocity fluctuations} (closed circles) and 
of {\it velocity derivatives} (open circles) measured in the DNS by Gotoh et al.\ 
at $R_\lambda = 380$ with the help of
the present theoretical PDF's $\hat{\itPi}^{(n)}(\xi_n)$ 
for {\it velocity fluctuations} (solid lines) and 
for {\it velocity derivatives} (dashed line) are plotted 
on (a) log and (b) linear scales.
The DNS data points are symmetrized by taking averages of 
the left and the right hand sides data.
The measuring distances, $r/\eta = \ell_n/\eta$, for the PDF of velocity fluctuations
are, from the second top to bottom: 2.38, 4.76, 9.52, 19.0, 38.1, 76.2, 152, 305, 
609, 1220. For the theoretical PDF's of velocity fluctuations, $\mu = 0.240$ ($q=0.391$), 
from the second top to bottom: $(n,\ \bar{n},\ q') =$
(20.7,\ 14.6,\ 1.60), (19.2,\ 13.1,\ 1.60), (16.2,\ 10.1,\ 1.58), 
(13.6,\ 7.54,\ 1.50), (11.5,\ 5.44,\ 1.45), (9.80,\ 3.74,\ 1.40), 
(9.00,\ 2.94,\ 1.35), (7.90,\ 1.84,\ 1.30), (7.00,\ 0.94,\ 1.25), 
(6.10,\ 0.04,\ 1.20), and
$
\xi_n^* = 1.10 \sim 1.43
$
($\alpha^* = 1.07$).
For the theoretical PDF of velocity derivatives, 
$(n,\ \bar{n},\ q')=(22.4,\ 16.3,\ 1.55)$, and
$
\xi_n^* = 1.06
$
($\alpha^* = 1.07$).
For better visibility, each PDF is shifted by $-1$ unit in (a) and 
by $-0.1$ in (b) along the vertical axis.
}}
\end{center}
\end{figure}
The dependence of the parameters $\alpha_0$, $X$ and $q$ on 
the intermittency exponent $\mu$ is determined, 
self-consistently, with the help of the three independent equations, i.e.,
the energy conservation:
$
\bra \epsilon_n/\epsilon \ket = 1
\label{cons of energy}
$ (equivalently, $\tau(1) = 0$),
the definition of the intermittency exponent $\mu$:
$
\bra \epsilon_n^2/\epsilon^2 \ket 
= \delta_n^{-\mu}
\label{def of mu}
$ (equivalently, $\mu=1+\tau(2)$),
and the scaling relation:
$
1/(1-q) = 1/\alpha_- - 1/\alpha_+
\label{scaling relation}
$
with $\alpha_\pm$ satisfying $f(\alpha_\pm) =0$. 
Here, $\epsilon$ is the energy input rate to the largest eddies.
The average $\bra \cdots \ket$ is taken with $P^{(n)}(\alpha)$.

The PDF's extracted by Gotoh et al.\ 
from their DNS data \cite{Gotoh02} at $R_\lambda = 380$ 
are shown, on log and linear scales, 
in Fig.~1 both for {\it velocity fluctuations} and 
for {\it velocity derivatives}, and are analyzed by the theoretical formulae
(\ref{PDF large}) and (\ref{PDF small}) for PDF's.
We found the value $\mu = 0.240$ by analyzing the measured scaling exponents 
$\zeta_m$ of velocity structure function with the formula given above,
which leads to the values $q = 0.391$, $\alpha_0 = 1.14$ and $X = 0.285$.
Through the analyses of the PDF's for velocity fluctuations in 
Fig.~\ref{velocity fluctuations log}, we extracted quite a few information of 
the system \cite{AA7,AA6,AA8,AA10,AA12}. 
Among them, we only quote here
the dependence of $q'$ on $r/\eta$:
$q'= -0.05 \log_2 (r/\eta) + 1.71$ \cite{AA12}.

\section{Granulence}

Let us now analyze the velocity fluctuations in glanulence 
simulated by Radjai and Roux \cite{Radjai02}. 
Since they observed that the fluctuations share the scaling characteristics of 
fluid turbulence, we try to investigate the system by means of MFA which
extracted, successfully, the rich information out of turbulence as was seen 
in the previous section. 
\begin{figure}[htb]
\begin{center}
\leavevmode
\epsfxsize=120mm
\epsfbox{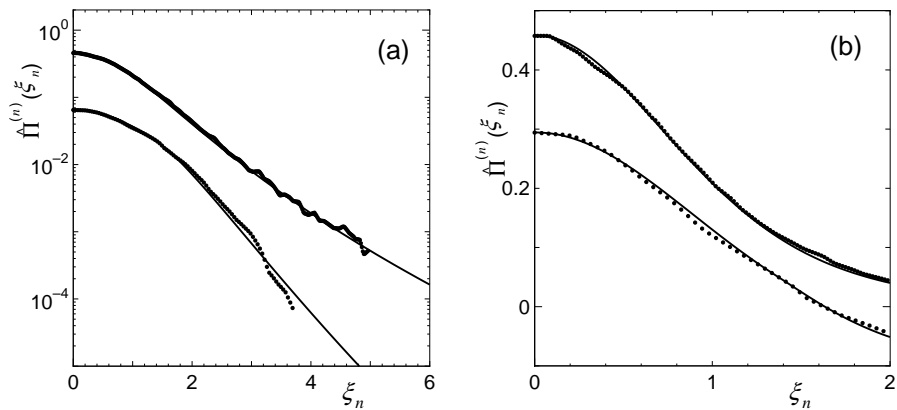}
\label{granulence}
\caption{{\small
Analysis of the experimental PDF of {\it fluctuating velocities}, 
measured in the quasistatic flow of granular media by Radjai and Roux,
with the help of the present theoretical PDF $\hat{\itPi}^{(n)}(\xi_n)$ 
for {\it velocity fluctuations} (solid lines) are plotted 
on (a) log and (b) linear scales.
The experimental data points are symmetrized by taking averages of 
the left and the right hand sides data.
The integration time $\tau$, normalized by a shear rate,
for the experimental PDF 
are, from top to bottom, $10^{-3}$, $10^{-1}$.
For the theoretical PDF, 
$\mu = 1.347$ ($q=0.930$), from top to bottom: $(n,\ q') =$
(88.0,\ 1.28), (40.0,\ 1.22), and
$
\xi_n^* = 1.14
$,
$
1.14
$
($\alpha^* = 0.364$).
For better visibility, each PDF is shifted by $-1$ unit in (a) and 
by $-0.1$ in (b) along the vertical axis.
}}
\end{center}
\end{figure}
The power spectrum of the fluctuating velocity field on one-dimensional cross sections
exhibits a clear power-law shape with the slope $-\beta$ 
with $\beta \approx 1.24$ \cite{Radjai02},
which is quite similar to the power-law behavior with the slope $-5/3$ in the inertial range 
of the Kolmogorov spectrum \cite{K41}. However, the granular model is an assembly of 
frictional disks, the power-law observed in 
granulence does not mean the energy conservation in contrast with the case of 
the energy cascade model for fluid turbulence.

For the conditions to determine the parameters $\alpha_0$, $X$ and $q$, 
we adopt, instead of the energy conservation, the slope of the power spectrum, i.e.,
$
\beta = 1 + \zeta_2 = 2-\tau(2/3)
$
in addition to the definition of the intermittency exponent and the scaling relation.
The latter two are the same as those for turbulence.
As there is no experimental data, for the present, to determine 
the intermittency exponent $\mu$ for granulence, we cannot have 
the values of the three parameters through the three conditions.
Therefore, we determine the value of the intermittency exponent by adjusting
the observed PDF with the theoretical formulae (\ref{PDF large}) and (\ref{PDF small}),
since the accuracy of the formulae in the analysis of PDF's for turbulence 
is quite high as was shown in the previous section.
The best fit of the observed PDF of fluctuating velocities
by the formulae (\ref{PDF large}) and (\ref{PDF small}) is shown
in Fig.~\ref{granulence}. We found the value $\mu = 1.347$ giving 
$q=0.930$, $\alpha_0 = 0.377$ and $X=0.050$.
By making use of the mass exponent with these values, we have 
$
\bra \epsilon_n/\epsilon \ket = \delta_n^{-\tau(1)}
$
with 
$
\tau(1) = 0.648
$
representing a breakdown of energy conservation.
It is attractive to see that the result is quite close to 
$
\bra \epsilon_n/\epsilon \ket =3/2
$
which may be consistent with the coefficient of friction 0.5 for 
the simulation \cite{Radjai02}.
We further extract the relation between $\tau$ and $\ell_n$ as
$
\tau = 1.3\ \delta_n^{\ 0.131}
$
by comparing the observed flatness and the one with the theoretical PDF's
(\ref{PDF large}) and (\ref{PDF small}).
This relation may be a manifestation of the fact that 
Taylor's frozen turbulence hypothesis does not work for granulence.

\section{Prospects}

We showed with the help of MFA 
that the system of turbulence and of granulence have, actually, common 
scaling feature in their velocity fluctuations as was pointed out 
by Radjai and Roux \cite{Radjai02}.
We expect that various observation of granulence will be reported 
at higher statistics, and that one can extract more information out of the data
to determine the underlying dynamics for granulence in the near future.



\begin{thebibliography}{10}

\bibitem{AA}
T.~Arimitsu and N.~Arimitsu.
\newblock {\em Phys. Rev. E}, 61:3237--3240, 2000.

\bibitem{AA1}
T.~Arimitsu and N.~Arimitsu.
\newblock {\em J. Phys. A: Math. Gen.}, 33:L235--L241, 2001.

\bibitem{AA10}
T.~Arimitsu and N.~Arimitsu.
\newblock cond-mat/0210274 2002.

\bibitem{AA12}
T.~Arimitsu and N.~Arimitsu.
\newblock { In \em Highlights in~Condensed Matter Physics (AIP Conference
  Proceedings~695)}, editor, A.~Avella, R.~Citro, C.~Noce and M.~Salerno.
  American Institute of Physics, 2003.

\bibitem{Radjai02}
F.~Radjiai and S.~Roux.
\newblock {\em Phys. Rev. Lett.}, 89:064302, 2002.

\bibitem{Tsallis88}
C.~Tsallis.
\newblock {\em J. Stat. Phys.}, 52:479--487, 1988.

\bibitem{Renyi}
A.~R\'{e}nyi.
\newblock {\em Proc.\ 4th Berkeley Symp.\ Maths.\ Stat.\ Prob.}, 1:547, 1961.

\bibitem{Havrda-Charvat}
J.H. Havrda and F.~Charvat.
\newblock {\em Kybernatica}, 3:30--35, 1967.

\bibitem{Frisch-Parisi83}
U.~Frisch and G.~Parisi.
\newblock { In \em Turbulence, Predictability in~Geophysical Fluid~Dynamics, and
  Climate Dynamics}, editors, M.~Ghil, R.~Benzi and G.~Parisi, page~84,
  New York, 1985. North-Holland.

\bibitem{Meneveau87b}
C.~Meneveau and K.R. Sreenivasan.
\newblock {\em Nucl. Phys. B}, 2:49--76, 1987.

\bibitem{AA3}
T.~Arimitsu and N.~Arimitsu.
\newblock {\em Prog.~Theor.~Phys.}, 105:355--360, 2001.

\bibitem{AA4}
T.~Arimitsu and N.~Arimitsu.
\newblock {\em Physica A}, 295:177--194, 2001.

\bibitem{Gotoh02}
D.~Fukayama T.~Gotoh and T.~Nakano.
\newblock {\em Phys. Fluids}, 14:1065--1081, 2002.

\bibitem{AA7}
T.~Arimitsu and N.~Arimitsu.
\newblock {\em J. Phys.: Condens. Matter}, 14:2237--2246, 2002.

\bibitem{AA6}
T.~Arimitsu and N.~Arimitsu.
\newblock {\em Physica A}, 305:218--226, 2002.

\bibitem{AA8}
N.~Arimitsu and T.~Arimitsu.
\newblock {\em Europhys. Lett.}, 60:60--65, 2002.

\bibitem{K41}
A.N. Kolmogorov.
\newblock {\em Dokl.\ Akad.\ Nauk SSSR}, 30:301--305, 1941.

\end{thebibliography}

\end{document}